Title: Photoinduced electrification of solids. III. Temperature dependences
Authors: O. Ivanov, Eugene Leyarovski,[†] V. Lovchinov, Chr. Popov, M. Kamenova, and M. Georgiev (Instirute of Solid State Physics, Bulgarian Academy of Sciences, Sofia, Bulgaria)
Comments: 10 pages of wording plus 14 figures, all pdf format
Subj-class: cond-mat

Two preceding parts of a paper (cond-mat/0508457, cond-mat/0508460) considered the heuristic values of recent experiments pointing to a nearly universal occurrence of photo voltages across solid surfaces under short-circuit conditions. These voltages arise by virtue of a variety of spectrally dependent mechanisms activated by incident photons. For the visible range, the photo voltages are obliged to the photo detachment of ions which leave the surface charged. In an attempt to learn more, we now study short-circuit photo voltages in well-defined materials including high-$T_c$ superconductors within a broad temperature range down to liquid nitrogen. We believe our data give a new insight into the process.

1. Introduction

For quite a bit of time now, experimentalists at these laboratories have obtained a good deal of evidence for the occurrence of short-circuit photovoltages on illumination of solid surfaces. The phenomenon is almost universal in that it occurs on any surface under light in the visible. Evidence has also accumulated to suggest that complementary mechanisms might be operative on both sides of the visible range, in the near infrared and in the ultraviolet as well [1].

All this came to show that there might be an universal mechanism operative within the wide spectral range where the quantum nature of the photons is the essential factor. We considered it likely that the common mechanism operating within the wider spectral range might well be the ionic processes triggered by an excitation of the electronic subsystem by the incident photons.

We also worked out an analysis based on the oscilloscopic observation of photovoltage transients making it possible to conclude in a relatively rapid way on the time constants involved in the photo charging process at given temperature [2]. But, we also believe that temperature itself is another agent which may be found useful for revealing the universal photo charging mechanism. Indeed, temperature dependent measurements can provide data on activated processes and therefrom on the barriers involved at various stages of the process. Such data will be presented and discussed below.

2. Experimental

The experimental arrangement has been described at length elsewhere [3]. For the temperature measurements, the condenser was placed in a vacuumed cryostat maintaining the temperature to within ±0.5 °C. The condenser box containing the sample was placed into a cryostat allowing scanning the temperature from 78 K to 300 K at a rate of 3 K/min on either cooling or warming. The accuracy of maintaining the temperature was 0.05 K.

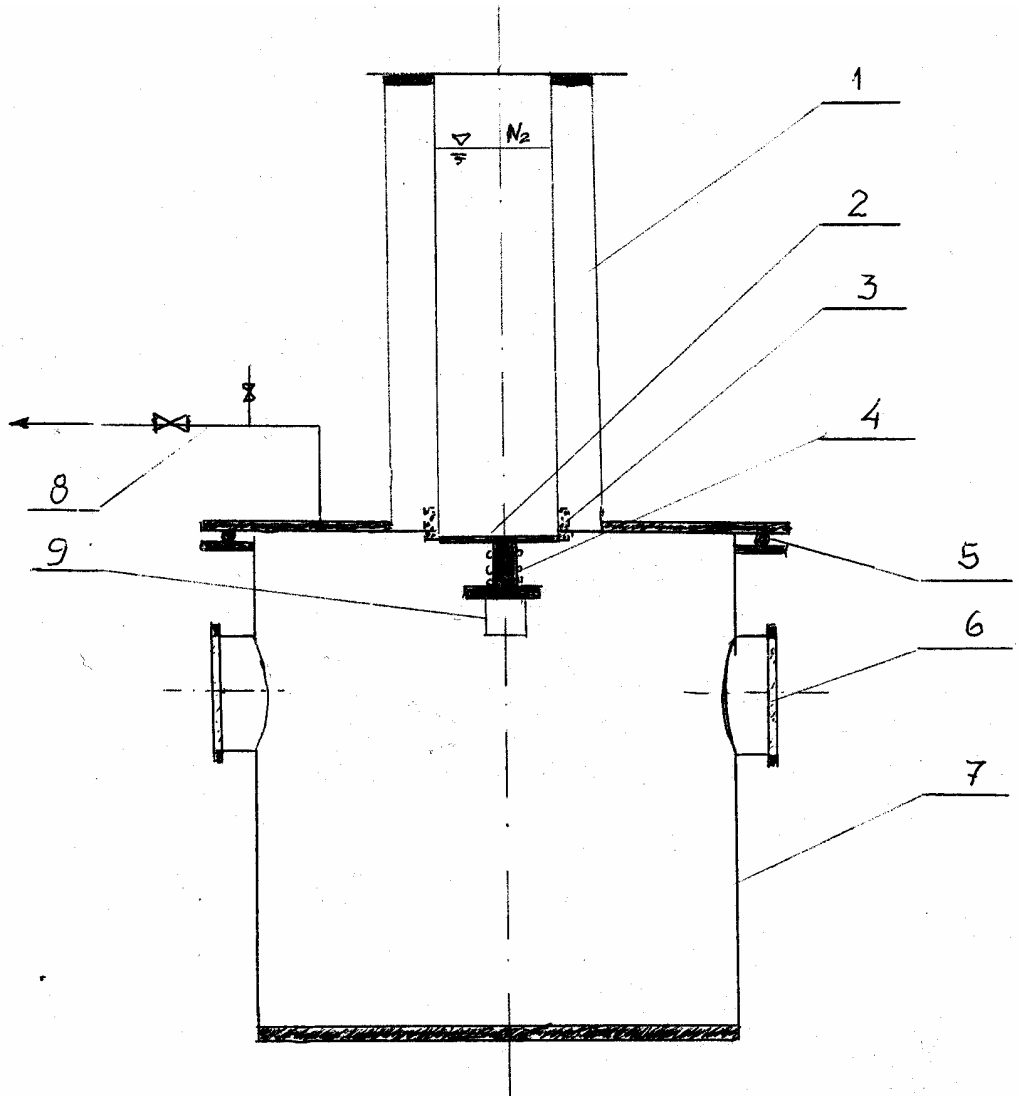

Figure 1

Schematic drawing of the cryostat used for temperature measurements. Notations: 1- Dewar, 2- copper bottom, 3- carbon, 4- heater, 5- rubber ring, 6- window, 7- operational volume, 8- vacuum line, 9- sample.

The excitation light wavelength was $\lambda = 488$ nm. The light chopping rate was $f = 180$ Hz. The disk chopper produced equal dark and light intervals. The diameter of the laser beam was d = 2 mm. Most of the experimental work was done under two laser powers: at P = 50 mW the output signal was V ~ 20 µV, at P = 280 mW the output was V ~ 200 µV. The schematic drawing of a cryostat used for photocharge measurements is shown in Figure 1.

### 3. Results

We investigated some key materials, such as a high-$T_c$ YBCO ceramics (in one instance $YBa_2Cu_3O_7$ with $T_c$ = 92 K), a nonsuperconducting ceramics ($YBa_2Cu_3O_6$?), as well as copper, aluminum and gold metals. The temperature range of the measurements was between 78 ÷300 K. Graphic results of the present study will be shown in Appendix.

#### 3.1. Ceramic materials

##### 3.1.1. Superconducting *YBa₂Cu₃O₇*

The superconducting ceramics exhibited kind of a hysteresis in that they followed a temperature path on warming not identical with the preceding one on cooling. Irrespective of the differences, both the temperature dependencies on warming and cooling exhibited similar temperature behavior: As a matter of fact, the signal increased in magnitude almost abruptly as the system was cooled down to 80 K. Before the abrupt change occurred, the signal was virtually insensitive to the temperature down to 100 K. This is shown in Figure 2 in a typical case. Alternatively on warming the system from 76 K upwards the signal was seen falling down much more slowly as the temperature was raised up to 200 K. The temperature-insensitive portion of the signal now extended between 200÷300 K. The observed hysteresis in *YBa₂Cu₃O₇* is seen in Figure 2. Some structures seem to appear in the temperature-insensitive range about 230 K and 280 K.

The superconducting $T_c$ was attained near the end of the cooling run but it is hard to point out any clear relationship between $T_c$ and the abrupt sensitization because of insufficient accuracy of the temperature measurement. What we can say with some degree of credibility is that the sensitization started out somewhere within the superconducting range on cooling and degraded beyond that range on warming the material. Accordingly, we are tempting to conclude that the experiments tell of a photosensitization somehow related to superconductivity. If so, the abrupt rise in photoelectric signal near $T_c$ could be used as a fingerprint for the superconductivity of a ceramic material.

##### 3.1.2. Nonsuperconducting NSCC ($YBa_2Cu_3O_6$ ?)

There are two cooling curves and one warming curve shown in Figure 3. The cooling curve exhibits the same general character as before except for a small dip near 100 K, not appearing in the superconducting ceramics. The cooling curve depicted by triangles show an even wider and better pronounced minimum again about 100 K. The warming curve by squares degrades to a minimum situated at about 220 K which is beyond any doubt. The non-superconducting ceramics NSCC undergoes a phase transition within the range of the minima. The one at the minimum near 220 K on warming may be related to the well-known transformation of tilting the $CuO_6$ octahedra. Another one at the minimum near 100 K on cooling is presently less than well understood.

### 3.1.3. Nonsuperconducting $PrBa_2Cu_3O_{7-\delta}$

The temperature dependence on cooling as in Figure 4 shows the signal gradually increasing as the temperature is lowered with steeper portion on both sides of a horizontal path.

### 3.1.4. Superconducting $Nd_{0.9}Pr_{0.1}Ba_2Cu_3O_7$

The material turns superconducting at $T_c \sim 57$ K. Despite the lower Tc the signal again exhibits an abrupt rise near 80 K. The temperature dependence of the photovoltage signal is shown in Figure 5.

### 3.1.5. Piezoceramics

The temperature dependence is shown in Figure 6 exhibiting a broad twin maximum about 160 K on cooling. It makes a convincing case giving credibility to the method.

## 3.2. Metals

### 3.2.1. Copper

Another convincing feature appears in the temperature dependence of the photocharging signal from Cu, as shown in Figure 7. The signal is now seen to pass through a maximum at 90 K on cooling and at about 180 K on warming. These peaks were not observed in the superconducting ceramics, though they may have appeared at temperatures beyond the measuring range which is less likely. Alternatively, the peaks may have been an inherent property of the metals and not of the superconducting ceramics. An overall reduction of signal with time was observed. The copper sample has not been subjected to any chemical cleaning before the photocharging experiment.

### 3.2.2. Gold

The photovoltage vs. temperature dependencies on cooling an Au sample in two separate runs are shown in Figure 8. Both exhibit a broad maximum near 200 K. The discrepancies show that the reproducibility along the photovoltage axis is not too high.

### 3.2.3. Aluminum

The temperature dependence of the photovoltage outcome from Al is shown in Figure 9 exhibiting a broad maximum on cooling, also near 200 K.

## 3.3. Elemental semiconductor (Si)

The photovoltage signal first increases smoothly on cooling the sample down to 90 K but then changes polarity abruptly at 80 K. The temperature dependence is shown in Figure 10

### 3.4. Natural mineral (Serpentine)

The temperature dependence of the photovoltage signal on cooling is composed of a sole maximum centered at 190 K, as shown in Figure 11.

### 3.5. Insulator (Textolyte)

The relevant temperature dependence of photovoltage signal plain down to about 100 K is seen to rise abruptly on further cooling, as in Figure 12. A rich structure can be seen down to 80 K.

### 3.6. Biomaterial (Wood)

The temperature dependence of the photovoltage signal obtained on cooling is remindful of what comes out from an isolator or even from a high-$T_c$ superconductor, as in Figure 13. Being somewhat less structured down to about 90 K, it rises abruptly on further cooling.

### 3.7. Binary compound (GaAs)

There are both temperature and intensity dependences taken on this material. In particular, the temperature dependence of the photovoltage on warming up the material shows voltage decreasing with temperature. Typical oscillograms showing the waveforms applied to photocharging measurements on n-GaAs are shown in Figure 14.

## 4. Discussion

At the chopping rate of 180 Hz the light-on intervals (equal to the light-offs) are 2.78 ms each. For the inherent time constants of the order of less than 1 ms the photocharging system will have time to saturate under light-on and to relax under light-off. As mentioned earlier, we considered the time-kinetics only without invoking the continuity equations which procedure albeit simpler may prove appropriate for describing surface conditions. Two operational modes are considered, as introduced in Part II.

### 4.1. Negative-U mode (U<0)

For the negative-U dihole mechanism, the saturated short-circuit conductivity under light-on is:

$$\sigma(\infty) / e\mu = p(\infty) + 2p_b(\infty) - 2p_d(\infty) \qquad (1)$$

where $\mu$ is the drift mobility assumed the same for all the quasiparticles, while the saturated quasiparticle densities are: $p(\infty)$ of the free holes, $p_b(\infty)$ of the diholes and $p_d(\infty)$ of the photodetached ions. Fields arising from gradients across the surface layers have been ignored for simplicity. We have solved for the saturated densities under two conditions, slow electron-hole recombination SEHR (exact solution) and fast electron-hole recombination FEHR (approximate solution), respectively. From Part II we obtain using the exact SEHR solution:

$$p(\infty) = \sqrt{(G/\gamma)}$$

$$p_d(\infty) = G(1 + \tau_b/\tau_d)^{-1}\{t - (1/\tau_d + 1/\tau_b)^{-1}\}$$

$$p_b(\infty) = G\{t - (1 + \tau_b/\tau_d)^{-1}[t - (1/\tau_d + 1/\tau_b)^{-1}]\} \tag{2}$$

Here $G = \eta\kappa I$ ($\eta$- ionization quantum yield, $\kappa$- absorption coefficient, I- light intensity) is the electron-hole generation rate, $\gamma$ is the hole trapping coefficient. For the approximate FEHR solution:

$$p(\infty) = \sqrt{(G/\gamma)}$$

$$p_d(\infty) = G\tau(1 + \tau_b/\tau_r)^{-1}$$

$$p_b(\infty) = G(1/\tau_r + 1/\tau_b)^{-1}\{(1 + \tau_b/\tau_r)^{-1}(\tau/\tau_d) + 1\} \tag{3}$$

with $\tau = \tau_d[1 - (1 + \tau_b/\tau_r)^{-1}]^{-1}$, $\tau_r$ is the electron-hole recombination time, $\tau_b$ is the dihole lifetime, $\tau_d$ is the detached ion lifetime before coming back to the surface. Both $\tau_b$ and $\tau_d$ may be expected to be strongly dependent on the temperature. (Note a misprint in $\tau$ just below equation (29) on page 7 of Part II.)

Of the quantities entering in G only $\eta$ is markedly temperature dependent, provided the photoexcitation lifts the hole to a higher-lying excited state before it is further ionized to the valence (conduction) band by virtue of thermal agitation. In this case [4]:

$$\eta(T) = [1 + A\exp(E_i/k_BT)]^{-1} \tag{4}$$

where A is a constant and $E_i$ is the ionization energy (T – absolute temperature, $k_B$ – Boltzmann's constant). At low temperature ($k_BT \ll E_i$), $\eta(T) \sim A^{-1}\exp(-E_i/k_BT)$, at high temperature ($k_BT \gg E_i$), though, $\eta(T) \sim 1$. It seems likely that the ascending branch of the V(T) curve in Cu on warming as the temperature is raised above 80 K is described by the $\eta(T)$ curve up to the peak signal at about 140 K whose position is material-dependent. The descending branch on raising the temperature after the peak signal may be due to the loss of free quasiparticles to trapping. The declining trend of signal proportional to $\tau_b$ reflects the temperature dependence of the bound-hole lifetime which is $\tau_b = B\exp(E_b/k_BT)$ where B is another constant. As most of the quasiparticles are immobilized, the signal gradually levels off to become temperature-independent.

### 4.2. Self-trapped exciton mode (STE)

We introduce p, $p_b$ and $p_d$ for the surface density of uniholes, self-trapped excitons and desorbed ions, respectively. G is the electron-hole generation rate, $\tau_r$ is as defined above, $\tau_b$ is the STE lifetime against decomposition, $\tau_d$ is the detached ion lifetime against readsorption. Again, fields arising from gradients across the surface layers will be ignored.

From Part II we derive the saturated SEHR solutions under light-on at $\tau_r = \infty$:

$p(\infty) = G \tau_s$

$p_d(\infty) = G\{ (1+\tau_b/\tau_d)^{-1}[t - (1/\tau_d+1/\tau_b)^{-1}] - (\tau_s / \tau_b)(1/\tau_d+1/\tau_b)^{-1} \}$

$p_b(\infty) = G\{ -(1+\tau_b/\tau_d)^{-1}[t - (1/\tau_d+1/\tau_b)^{-1}] + (\tau_s / \tau_b)(1/\tau_d+1/\tau_b)^{-1} + (t - \tau_s) \},$   (5)

as well as the FEHR solutions at finite $\tau_r$:

$p(\infty) = G \tau_s$

$p_d(\infty) = G \tau (1 + \tau_b / \tau_r)^{-1}$

$p_b(\infty) = G (1/\tau_r + 1/\tau_b)^{-1} \{ (1 + \tau_b / \tau_r)^{-1} (\tau / \tau_d) + 1\}$   (6)

where $\tau = \tau_d [1 - (1 + \tau_b / \tau_r)^{-1}]^{-1}$, $\tau_s = [\gamma(p,T) n]^{-1}$. (See Part II for details.)

The conductivity equation holds good formally for the STE mode too:

$\sigma(t) = e \mu_p p(t) + 2e \mu_b p_b(t) - 2e \mu_d p_d(t)$   (7)

### 4.3. Implications

It is amazing how most of the materials investigated exhibit the abrupt increase in photo voltage as the sample is cooled down towards the nitrogen boiling point (LNT). Exceptions may be found in Figures 4, 7 and 11 which almost invalidates the artifact. The increasing branch rather reminds of the higher-temperature branch of a thermostimulated peak (TSP), even though it appears on warming up rather than on cooling down and may be indicative of the filling in of carrier traps. As stated below, we derived an activation energy of about 50 meV from the branch in Figure 2. A likely hypothesis supported by an almost universal occurrence is that the abrupt branch arises from a superficial process prerequisite to the ionic detachment which forms the basis for our suggested photo charging mechanism, the photodesorption. On the other hand, the estimated surface energy the order of 50 meV appears more or less typical of solids [5]. It also appears likely that the TSP's of Pr-Ba-Cu-O ceramics, Cu metal, and ammonite in Figures 4, 7, and 11, respectively, may be displaced to higher temperatures which implies higher surface adhesion energies.

Following earlier suggestions [1,2], free holes are photogenerated with a growing efficiency within the colder ascending branch of the photovoltage vs. temperature dependence on warming up the sample. They are by all means effectively trapped within the subsequent descending branch. It would not be astonishing if the ascending branch had not been observed in the superconducting material for it might have remained hidden beyond the measuring range. Alternatively, no evidence for a pair photogeneration may have been found for the simple reason that holes may have quickly paired, while majority pairs (diholes or hole Cooper pairs) have already been there in vast numbers so that the photoexcitation may have little to add. As a matter of fact, when the preexisting holes are too few, the photoexcitation is known leading to a photoinduced superconductivity. In high-$T_c$ materials with preformed dihole pairs the photoexcitation might eventually elevate the detachment of ions from the crystal thereby boosting the third (negative) component of the short-circuit current density [2].

At the same time, comparing with the saturated densities in Sections 4.1 and 4.2, we see that these densities are mostly limited by the lifetime $\tau = \tau_d [1 - (1 + \tau_b / \tau_r)^{-1}]^{-1}$. Considering the two groups of FEHR equations (3) and (6), we distinguish between two temperature ranges, high ($\tau_b \ll \tau_r$) and low ($\tau_b \gg \tau_r$), the relative temperature insensitivity of $\tau_r$ being the reference factor. In the former high T range $\tau \sim (\tau_d / \tau_b)\tau_r$ while in the latter low T range $\tau \sim \tau_d$. It follows that the photovoltage signal goes relatively flat if $\tau_d \sim \tau_b$ within the high temperature range and rises steeply $\propto \tau_d$ as the temperature is further lowered on cooling. Indeed, this is actually observed (see Figures 1÷5). The respective saturated densities meet $p_d(\infty) \sim G \tau_r (\tau_d / \tau_b) \geq p_b(\infty) \sim G \tau_r$ in both temperature ranges.

Given a superconducting material, it will be in a Bose condensed state at $T \leq T_c$, as defined by the formation of hole Cooper pairs. If photoholes couple to form diholes they may eventually fall in the condensed pair state which is the lowest-energy free particle state of the material. However, the lowest-energy particle state will be that of diholes trapped at (active) bonds. It follows that diholes will automatically go to the trapped state at an increasing rate as the temperature is lowered below $T_c$. Ultimately, the resulting trapped dihole density will tend to increase with lowering the temperature $\propto \tau_{TDS}(T)$, where $\tau_{TDS}(T) \propto \exp(E_{TDS} /k_BT)$ is the temperature-dependent lifetime of the trapped dihole state. (Indeed, from $dp_b/dt = G - p_b / \tau_{TDS}$ we get $p_b(\infty) = G\tau_{TDS}$.) Here $E_{TDS}$ is the thermal trap depth of a dihole immobilized at a bond. From the data in Figure 2 we estimate $E_{TDS} \sim 50$ meV as the activation energy of the abruptly rising branch on cooling YBCO near $T_c$.

However, the essential contribution to $\tau$ comes from the lifetime $\tau_d$ of the detached ion cloud, for both U<0 and STE. In so far as $\tau_d$ controls to a large extent the surface charge and thereby the equilibrium density of bulk charges, it should also control the equilibrium condensed state of the bulk pairs and possibly the free diholes. As a matter of fact, this assignment confirms that the desorbed ion concentration $p_d$ controls the electrostatic events at and beneath the surface within the measuring range about 80 K.

From equations (3) and (6) for the FEHR regime of both the U<0 and STE modes, the respective saturated densities are similar though not identical. For further considerations the FEHR regime is preferred because of experimental observations of the ultimate role of the electron-hole recombination in clearing the photocharge during the dark periods. The lifetimes can reasonably be expected to arrange numerically in the following order: $\tau_d \geq \tau_b > \tau_r \sim \tau_s$. From photovoltage transients shown elsewhere we deduce in a particular case $\tau_b \sim 1$ ms, $\tau_d \sim 3$ ms, $\tau_r \sim 0.1$ ms [6]. The photogeneration rate G enters as a scaling factor.

Going back to the cooling down run in superconducting ceramics (YBCO), we again focus on the abrupt rise of photovoltage near $T_c = 90$ K. It reminds of a boom of photocharges at a nearly constant temperature somehow related to superconductivity. Certainly, there have been early suggestions of an element common for laser sputtering and high-$T_c$ superconductivity which is the hole bipolaron [7]. At any rate it may be expected to form at $T_c$ or above. With hole bipolarons flooding the subsurface region, the rate of diholes appearing at negative-U sites will increase greatly and so will the photovoltage. We conclude that photocharging centers are by all means formed at negative-U sites when diholes get trapped thereat: Diholes residing at valent bonds will tend to break these bonds giving rise to ions or atoms desorbing from the surface. As the amount of desorbed ions increases so will the electric flow out of the surface. The photovoltage declining as the temperature is raised on warming may be indicative of the trapped dihole lifetime against desorption or the desorbed ion lifetime against re-adsorption. In either case the warming-up run provides data on the respective temperature-dependent lifetimes.

To summarize, we propose the following common steps leading to a photovoltage on warming up within the temperature range between 78÷300 K, as in Figure 7 pertaining to copper: At the lowest temperatures the signal increases with T as a result of the increasing ionization quantum yield (4) producing free charge carriers. The free holes rapidly combine to form diholes at negative-U sites. (In high-$T_c$ superconducting materials at $T \leq T_c$ the hole pairs are there and the role of light might also be to stimulate extra photoholes to negative-U sites, perhaps by photostimulated migration.) Passing through a material-dependent maximum as the yield saturates, the voltage then degrades down to low values as a result of the increasing number of diholes breaking bonds thereby giving rise to ions desorbed from the crystalline surface. The desorption rate increases with T which is understandable in view of the surface reactions [5]. As the outgoing ions are recharged at the electrodes, they come back to the surface to readsorb. The incoming ions produce a reverse flow which eventually reduces and even overcompensates the net short-circuit current recharging the surface. This can be evidenced in some instances as a flat dip after which the signal starts increasing as the temperature is further raised towards 300 K [6].

It is amazing how metals (Cu) behave in a way similar to poor conductors in many ways which perhaps comes to emphasize the common features of the phenomenon. The low flat yield on Cu warming around 80 K may be due to residual tunneling ionization, as found elsewhere though in another context [8]. We do not yet know much of the nature of local centers in metals (Cu) which behave like ones in semiconductors or insulators though the present experiments reveal similarities alongside with differences. Perhaps local surface centers originating from adatoms provide an explanation. Alternatively, it might be that the observed features are inherent of CuO rather than of the pure metal. Similar arguments may be raised with respect to the photovoltage signal from Al which is known to form a thin sapphire layer which keeps it from corrosion. Unlike Cu and Al, gold Au should behave like a genuine metal and indeed it exhibits a clear-cut structure peaked at ~ 200 K.

### 4.4. Hysteresis

In principle, different shapes of a temperature dependence of the photovoltage on cooling ($\partial T/\partial t < 0$) and warming ($\partial T/\partial t > 0$) could result from different rates $\partial T/\partial t$ of varying the temperature T with time t as in $T - T_0 = (\partial T/\partial t)(t - t_0)$. Indeed, if the temperature dependence is determined by some lifetime $\tau(T) = \tau_\infty \exp(\varepsilon/k_B T)$, which is the measurable quantity, then for $(\partial T/\partial t)' \neq (\partial T/\partial t)''$ we get $T' \neq T''$ and therefore $\tau(T') \neq \tau(T'')$. Clearly, the temperature mismatch will be the less significant the shorter the lifetime, so that unintentional differences of temperature vs. time rates will be more effective at the lower temperatures. As a matter of fact, we see the hysteresis spreading up to ~ 200 K or lower in the foregoing Figures 2 and 3. This can be regarded as a grounds for attributing the apparent hysteresis in photocharging to temperature artifacts.

Apart from the artifacts, there may be some additional reasons for the occurrence of a hysteresis in photocharging phenomena. Indeed, if photocharging centers are being formed on cooling, such as diholes resting on active bonds, most of the diholes will break their harboring bonds giving rise to desorbing ions. Nevertheless, some trapped diholes may survive the desorption trends to give a residual photocharge resulting in a hysteresis.

### 5. Conclusion

We believe to have manifested temperature dependences of the photocharging signal in selected materials among them high-Tc superconducting ceramics, non-superconducting ceramics, as well as copper, gold, and aluminum metal samples. All the three groups have common features such as the increase of signal on cooling down the material and the decrease of signal on warming it up. Most specifically, the temperature dependence of the signal on cooling YBCO showed an unusual abrupt rise about $T_c$ which seems inherent to the material undergoing a transition into the pair condensed state.

Our model considerations of the photocharging of solids due to photodesorption point to an ionic hallo being created around the surface which leaves the bulk of the material electrified. In a superconducting material the hallo controls the details of the equilibrium pair condensed state, such as the attaining of a thermodynamic equilibrium at given temperature near $T_c$. The signal rises abruptly because it is thermally activated around $T_c$ with the photodecomposition energy of bound dihole pairs. This explains to a large extent the observed shape of the photoresponse about $T_c$. From this point of view, the details of the abrupt voltage rise might be inherent of the superconducting materials and utilized as a practical means for a contactless detection of superconductivity by cooling down the material and monitoring its photocharging response.

## Acknowledgement

MG is grateful to Dr A Marshall Stoneham for an illuminating discussion on the role of pre-existing hole pairs in the laser sputtering of high-$T_c$ superconductor surfaces.

## Appendix
## Results represented graphically

Most of the data obtained so far are represented in a graphical form where the measured photovoltages V (U) are plotted versus the temperature T of the sample. In only a few cases are the results described quantitatively instead. We have uncovered a number of unexpected features, such as the structured plateau above certain temperature and the occurrence of well-defined voltage peaks superimposed on the plateau. It is hoped that these data will stimulate further fundamental research in a field often regarded as marginal. Indeed, over the years we became aware of some of our results being observed by others too but considered unwanted as "parasite µV photovoltages." Instead, we expect that some of these photovoltages may in fact provide the key to novel physics, now as the main physical facts in the normal voltage range have been uncovered and explained. Accordingly, we expect the photocharge research to continue, despite skepticism, under better sponsored conditions to fill in the gap between our present understanding of the range of small nanoparticle size and the small-signals coming from normal-size particles.

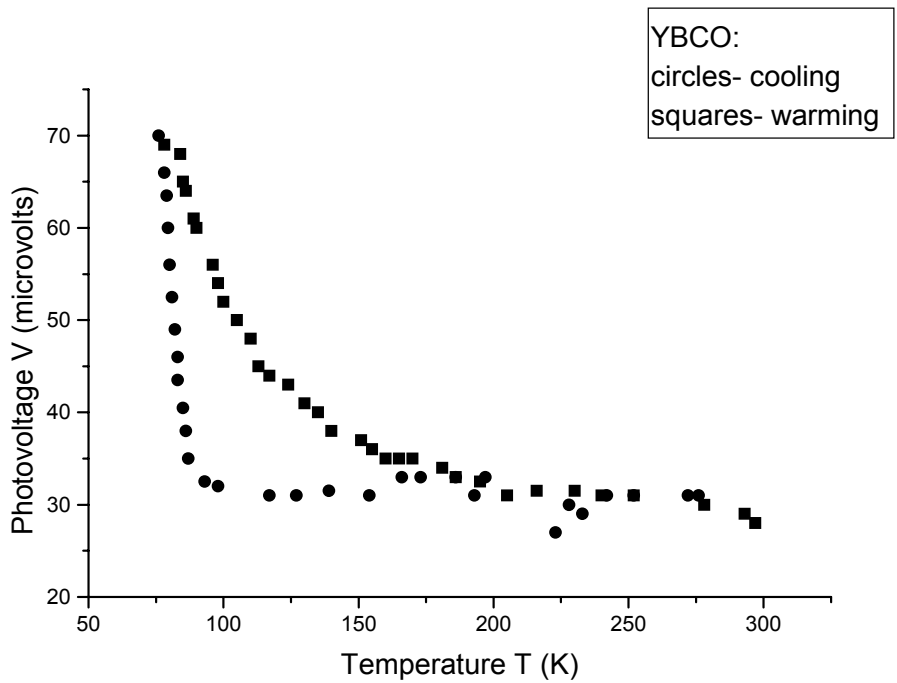

Figure 2

Temperature dependence of the photovoltage signal on cooling and subsequent warming of a superconducting YBCO ceramics ($T_c$ = 92 K). The cooling-down dependence is nearly flat down to about the liquid nitrogen temperature (LNT), then it rises abruptly on further cooling. The warming-up dependence shapes a similar trend though it is more gradual. The two exhibit a hysteresis below 200 K but are indistinguishable at higher temperatures.

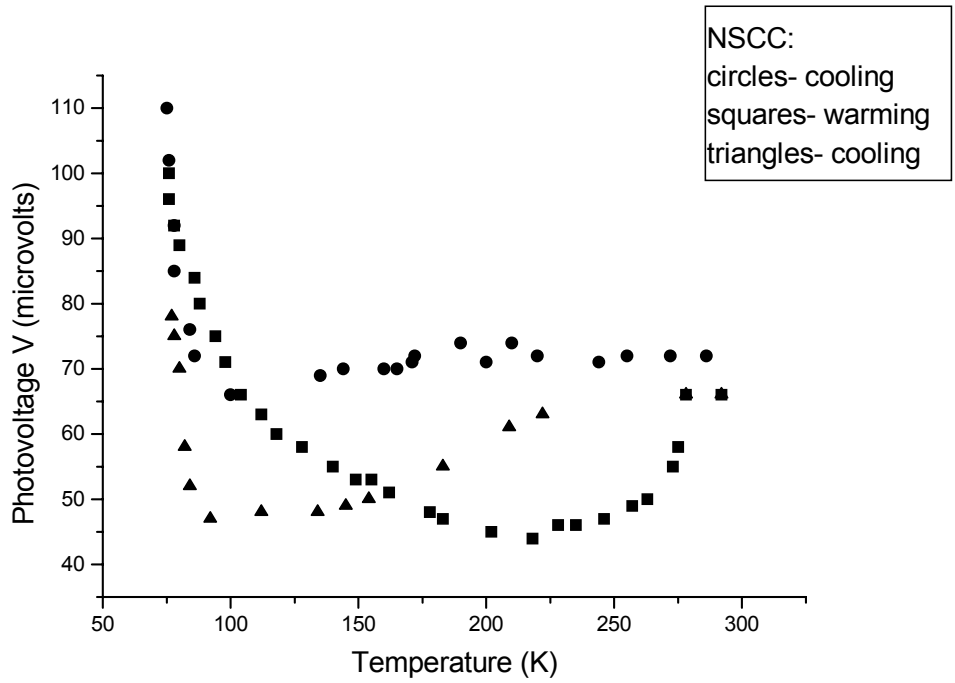

Figure 3

Showing two cooling-down runs and one warming-up run on a nonsuperconducting ceramics. The material is seen to undergo structural transformations at 110 K (cooling) and 220 K (warming), as shown by the respective minimuma on a pair of cooling-warming runs.

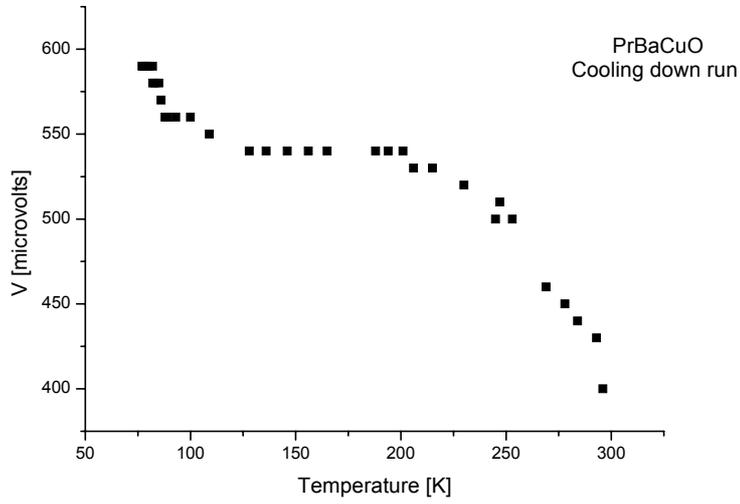

Figure 4

The cooling run of a nonsuperconducting $PrBa_2Cu_3O_{7-\delta}$ ceramics. Sometimes solid lines will be drawn to guide the eye.

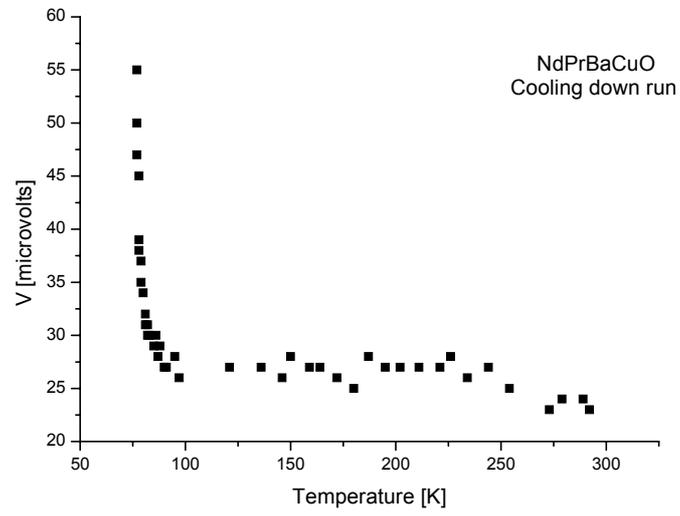

Figure 5

The cooling run on a superconducting $Nd_{0.9}Pr_{0.1}Ba_2Cu_3O_7$ ceramic sample ($T_c$ = 57 K). The plateau above 100 K seems to be structure rich.

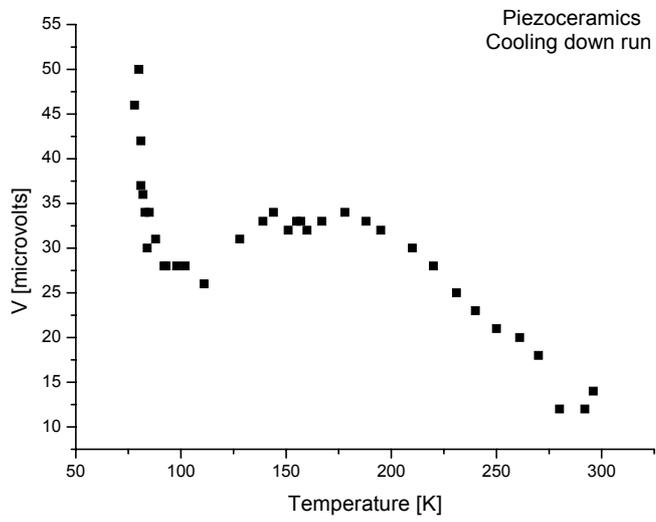

Figure 6

The photoresponse on cooling down a piezoceramic sample. The twin peaks near 150 K may signify a parallel behavior.

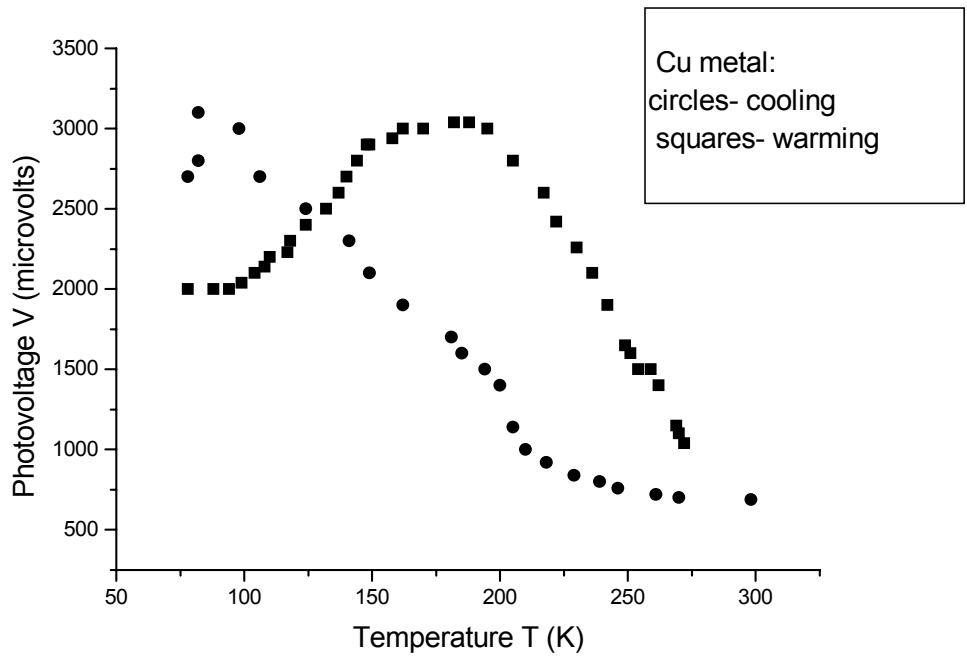

Figure 7

Showing the temperature dependencies of the voltage signal on photoelectrifying a copper metal plate. The cooling-down and warming-up dependencies, again exhibiting a hysteresis, are rich in features, as discussed in the text. See legend to insert

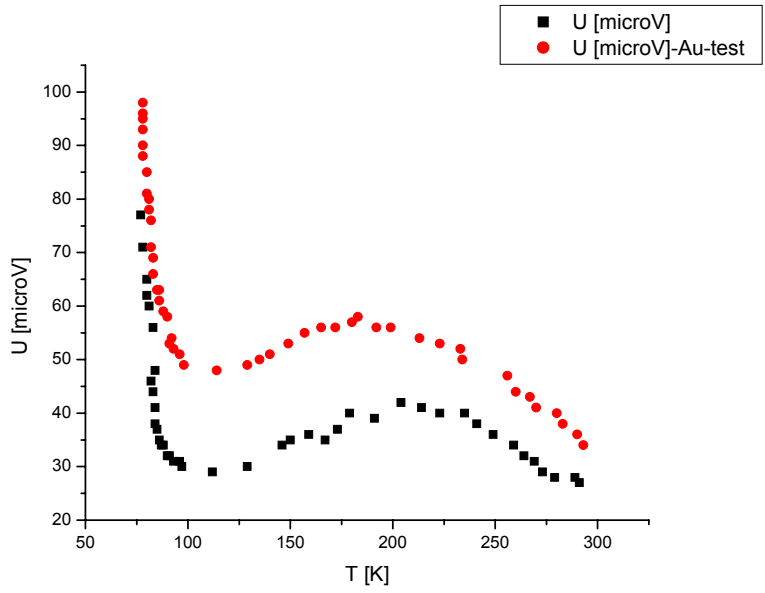

Figure 8

Two cooling down runs on a gold sample. The reproducibility along the temperature not voltage axis is fairly satisfactory.

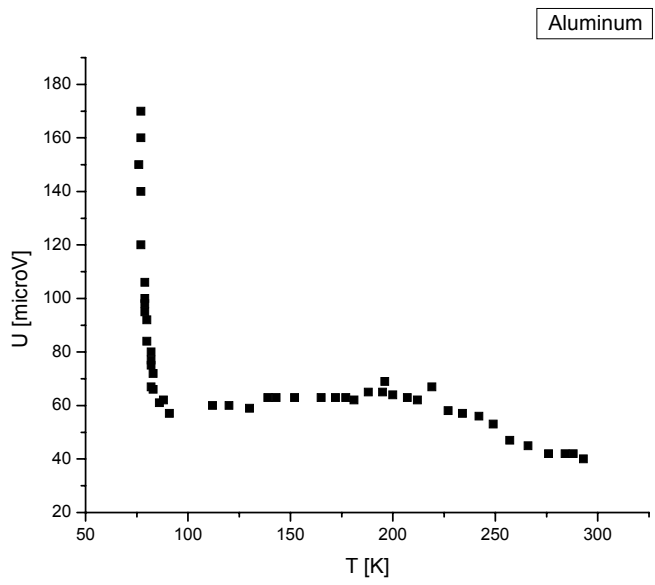

Figure 9

The cooling down run on an aluminum sample. There is some element common for the metallic character.

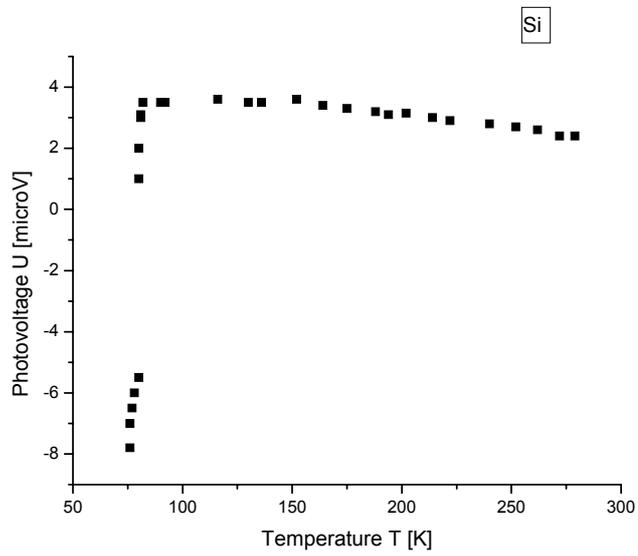

Figure 10

The cooling photo response of a Si sample. Unlike the remaining samples shown, silicon changes polarity of photo charging near LNT.

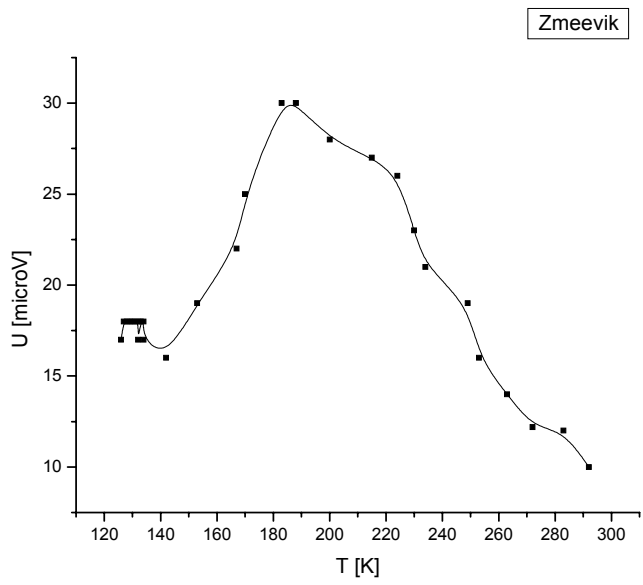

Figure 11

The cooling down run on a mineral: serpentine. The peak near 180 K reminds of the thermostimulated occurrences, though in the opposite direction.

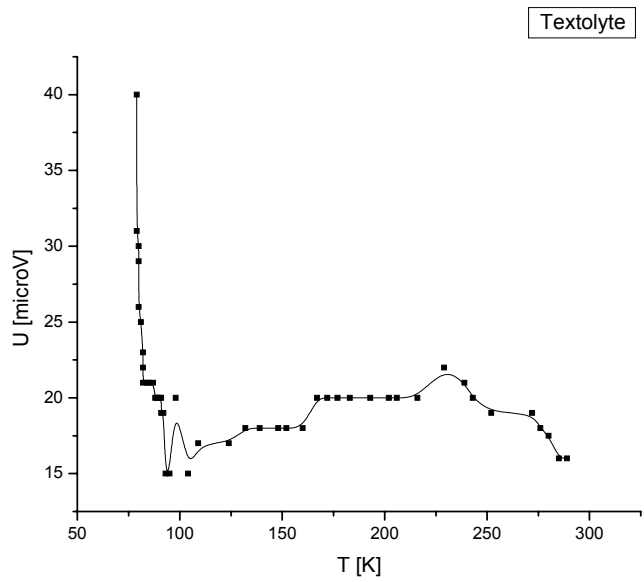

Figure 12

The cooling down run on an insulator (textolyte). The plateau is structure rich.

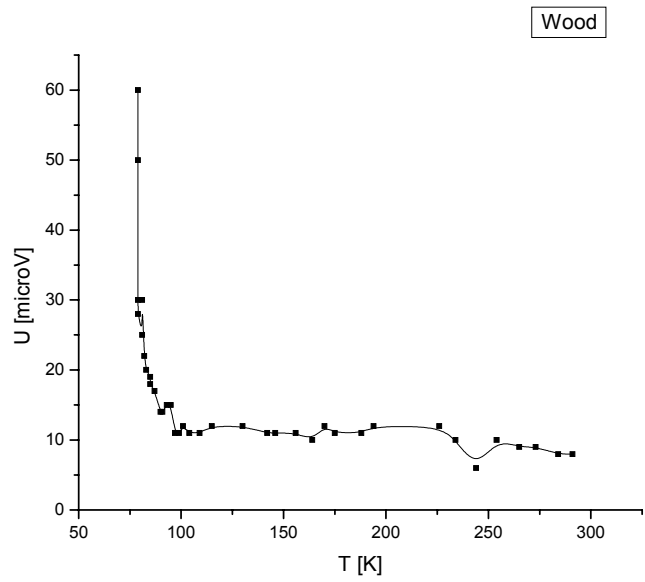

Figure 13

The cooling down run of a wooden sample. Again, the plateau is structure rich. The wood may be taken as an example for a biological sample.

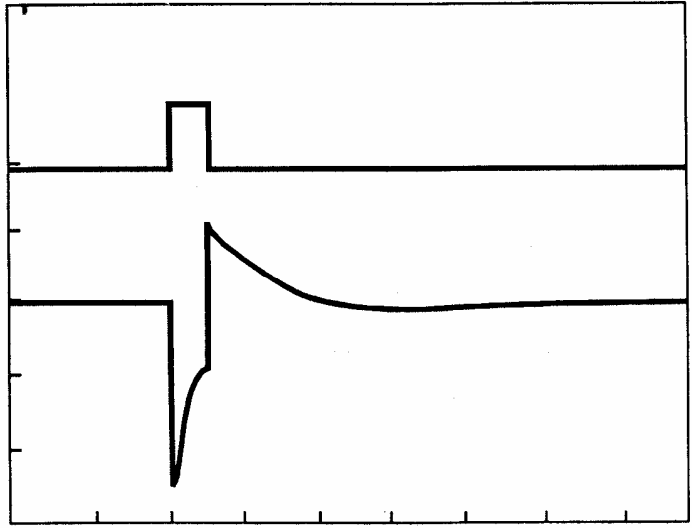

Figure 14

Oscilloscope waveforms obtained in a typical experiment on n-GaAs. The upper pattern is the light pulse, the lower one is the photovoltage response to the excitation. From the voltage vs. time dependencies under light-on and –off one obtains the relaxation times controlling the process. One division along the time axis amounts to 5 ms.